\documentclass[preprint]{article}
\usepackage[latin1]{inputenc}
\usepackage{amsmath}
\usepackage{amsfonts}
\usepackage{amssymb}
\usepackage{lpic}
\usepackage{natbib}
\usepackage{graphicx}

\begin{document}

\title{Internal Gerstner waves: applications to dead water}
\author{Raphael Stuhlmeier}
\date{}

\maketitle

\begin{abstract}
We give an explicit solution describing internal waves with a still water surface, modelling the dead water phenomenon, on the basis of the Gerstner wave solution to the Euler equations. 
\end{abstract}

\section{Introduction}

The phenomenon of \emph{dead water} was first investigated by Vagn Walfrid Ekman at the initiative of the Norwegian oceanographer and explorer Fridtjof Nansen, who encountered  it on the voyage of the \emph{Fram} through the Arctic Ocean, and subsequently proposed its study to Ekman's teacher Vilhelm Bjerknes.  Descriptions of dead water as encountered by early seafarers go back to antiquity, an engaging account of which is given by Ekman in \emph{The Norwegian North Polar Expedition} \citep{Nansen1906}.


 Dead water is so-called for its ability to slacken the speed of a vessel quite suddenly, and with no apparent cause  on the water surface. This phenomenon is prevalent when a layer of less dense water overlays a layer of denser water, due either to differences in salinity or temperature. Such instances of fresh or brackish water resting upon heavier sea-water occur at the mouths of rivers, particularly in the Norwegian fjords, where relatively little mixing of the waters takes place.  Internal waves generated at the interface between the two fluids may impede the progress and steerage of ships. 
 
 Causes and countermeasures for dead water reported by early mariners were manifold and at times quite amusing. It was believed that the dead water could be loosened from a ship by beating the water with oars or pouring petroleum into the water ahead of the ship. The belief that it was caused by a small \emph{remora} fish sucking fast to the ship's body was also widespread, an explanation tracing back to Pliny the Naturalist, and there abound stories of submarine magnetic rocks, or supernatural forces.

In the tradition of the early $20^{th}$ century, Ekman's work relied on a linearized theory (based on the discussion of internal waves by G.\ G.\  \citet{Stokes1847}), as did Lamb's article on the subject a decade later \citep{doi:10.1080/14786440408635511}. More recently, however, both theoretical and experimental investigations have highlighted the importance of nonlinear effects in connection with the dead water phenomenon. Mercier, Vasseur and Dauxois' experimental study \citep{Mercier2011} revisits Ekman's work and goes on to study interfacial waves in three-layer as well as linearly stratified fluids with a pycnocline. Motivated in part by this experimental study, 
 \citet{0951-7715-24-8-008} has given a sophisticated analysis of some nonlinear models for the dead-water phenomenon with the assumption that the wavelengths of the internal waves are long compared to the layer depths, while also incorporating aspects of the ship motion. We refer the reader also to references therein for a fuller account of the existing literature.

We will provide an explicit solution to the nonlinear governing equations for water waves describing such internal waves which leave no trace upon the surface of the water. While the passage of a ship will necessarily generate waves both on the water surface and at the pycnocline, the surface waves will be of considerably diminished amplitude \cite{Nansen1906}, and these long and low waves may well escape notice both by mariners as well as in the laboratory \cite[p.\ 194]{Mercier2011}. 

The mechanism of generation of these waves by a vessel is, regrettably, outside of the scope of this study. Hence we must content ourselves with a description of the wave motion itself, based on the simple insight that the classical Gerstner wave can propagate also at an interface between two liquid regions. The difference between the problem presented at the two-fluid interface and the usual water-wave problem at the free surface manifests in the fact that the pressure at the interface is no longer constant, but hydrostatic. The Gerstner wave construction allows the pressures at the interface to be matched, which introduces a reduced gravity into the problem.

\section{Governing equations}

The physical situation as seen in figure 1 may be captured as follows:
Let our fluid be contained in the region $\{(x,y) \mid x \in \mathbb{R}, \, y\leq 0\},$ while the upper half-plane $y> 0$ is assumed to be composed of air with negligible density which does not interact with the fluid below. We will assume that the upper layer of fluid has a constant density $\rho_1,$ and the lower fluid a density $\rho_2,$ with $\rho_1 < \rho_2,$ the interface between these immiscible layers being denoted by $\eta(x,t).$

\begin{figure}[h]
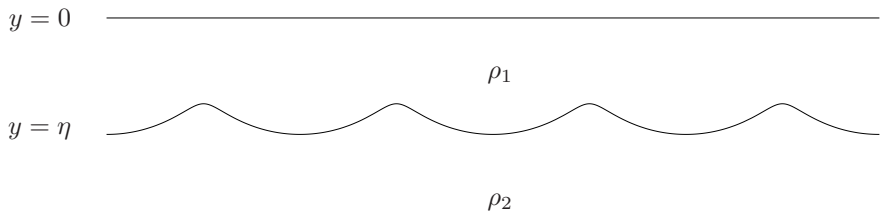

\centering
\begin{lpic}[l(10mm),r(3mm),t(5mm),b(10mm)]{deadwater(0.68)}
\lbl[t]{80,15;$\rho_1$}
\lbl[t]{80,-10;$\rho_2$}
\lbl[m]{-10,24;$y=0$}
\lbl[t]{-10,4;$y=\eta$}
\end{lpic}
\caption{Internal waves at the interface between lighter (density $\rho_1$) and heavier (density $\rho_2$) water.}
\end{figure}

The governing equations for two-dimensional, incompressible, inviscid free surface flow are the equation of mass conservation
\begin{equation}
\label{masscons}
u_{x} + v_{y} = 0
\end{equation}
and the Euler equation
\begin{equation}
\label{Euler}
u_t + uu_x + vu_y = - \frac{P_x}{\rho}, \quad v_t + uv_x + vv_y = -\frac{P_y}{\rho} - g,
\end{equation}
supplemented by suitable boundary conditions cf.\ \citep{Constantin2012d}. Here the horizontal and vertical components of the velocity field are $u(x,y,t)$ respectively $v(x,y,t),$ while $P(x,y,t)$ denotes the pressure, and $g$ is the constant gravitational acceleration. Subscripts indicate a partial derivative.

In order to decouple the motion of the air from that of the water, we introduce a dynamic boundary condition at the water surface:
\begin{equation} 
\label{DSBC}
P = P_{atm} \text{ on } y = 0,
\end{equation}
which specifies that the pressure at the upper interface is equal to the atmospheric pressure $P_{atm}.$
We are interested in an interface describing steady waves, and may thus assume $\eta = \eta(x+ct).$
In order to assure that this interface separates the fluids completely, one introduces a further, kinematic, boundary condition specifying that the fluid velocity thereon is wholly tangential. Temporarily denoting by $(u^1,v^1)$ and $(u^2,v^2)$ the velocity fields of the upper respectively lower layer, this condition takes the form
\begin{equation}
\label{SKBC} 
v^i = \eta'(u^i + c) \text{ on } y = \eta, \text{ for } i=1,2.
\end{equation}
As we shall not account for motion far below the water surface, there is no need to specify a similar condition for the sea-floor. Instead, we will assume that our motion dies out at great depth
\begin{equation} \nonumber
\label{velocity_decay}
u^2,v^2 \rightarrow 0 \text{ as } y \rightarrow -\infty.
\end{equation}

The top layer of water will be assumed to be moving with the wave at speed $c$ and the surface of the water taken to be flat, such that $(u^1,v^1)=(-c,0).$ This satisfies the kinematic boundary condition at the interface \eqref{SKBC}, and, trivially, the same condition at the water surface.
Hence, in the near-surface layer, \eqref{Euler} becomes
\begin{equation}
\label{Euler_stillwater} \nonumber
P_x = 0 \text{ and } -P_y = \rho_1 g
\end{equation}
which, along with the condition \eqref{DSBC} that the pressure be equal to the atmospheric pressure at the flat water surface, yields
\begin{equation}
\label{pressure_top}
P = P_{atm} - \rho_1 g y
\end{equation}
throughout the upper fluid layer.

\section{Structure of the solution}
In order to describe the internal wave we are interested in, we specify the particle paths of our solution in Lagrangian, or material, coordinates $(a,b),$ with $a \in \mathbb{R}$ and $b\leq b_0 \leq 0$, where $b_0$ will describe the interface between the two fluids. Indeed, the wave motion is simply that of Gerstner's trochoidal solution \citep{ANDP:ANDP18090320808}, though we shall see that the pressure at the interface $\eta$ will lead to some modifications. The particle trajectories are given by
\begin{align}  \nonumber
x = a + \frac{e^{mb}}{m} \sin m(a+ct), \\  \nonumber
y = b - \frac{e^{mb}}{m} \cos m(a+ct).
\end{align}
Then the horizontal and vertical velocities are 
\begin{align}  \nonumber
u &= c e^{mb} \cos m(a+ct), \\  \nonumber
v &= c e^{mb} \sin m(a+ct),
\end{align}
and we find the accelerations to be:
\begin{align}  \nonumber
\frac{Du}{Dt} &= -c^2 m e^{mb} \sin m(a+ct), \\  \nonumber
\frac{Dv}{Dt} &= c^2 m e^{mb} \cos m(a+ct).
\end{align}
Because the motion is two-dimensional, we may also specify the velocity field via a stream function,
\begin{equation} \label{streamfunction}  \nonumber
\psi = c\left(b - \frac{e^{2mb}}{2m}\right), 
\end{equation}
whereupon we may identify the streamlines to be curves of constant $b.$
Setting for readability $m(a+ct) = \theta$ and $mb = \xi,$ we find
\begin{align}  \nonumber
P_x &= \rho_2 c^2 m e^{\xi} \sin \theta, \\  \nonumber
P_y &= - \rho_2(c^2 m e^{\xi} \cos \theta + g).
\end{align}
Now, transforming to material coordinates via 
\begin{equation} \label{coord_transform}
\begin{pmatrix}
P_a \\
P_b
\end{pmatrix}
=
\begin{pmatrix}
1 + e^\xi \cos \theta & e^\xi \sin \theta \\
e^\xi \sin \theta & 1-e^\xi \cos \theta
\end{pmatrix}
\begin{pmatrix}
P_x \\
P_y
\end{pmatrix}
\end{equation}
yields
\begin{align}  \nonumber
P_a &= \rho_2 (c^2 m - g) e^{\xi} \sin \theta,  \\  \nonumber
P_b &= \rho_2 c^2 m e^{2\xi} - \rho_2 g + \rho_2 (g - c^2m) e^\xi \cos \theta, 
\end{align}
which we may integrate to yield a pressure
\begin{align}
\label{pressure_gerstner}
P = -\rho_2 (c^2 m - g) \frac{e^{\xi}}{m} \cos \theta + \rho_2 \frac{c^2}{2} e^{2\xi} - \rho_2 gb + C.
\end{align}
At the interface, which we take as the streamline $b = b_0$, the above pressure must match that for the still fluid \eqref{pressure_top}, given in material coordinates by
\begin{equation}
\label{pressure_still}  
P = P_{atm} - \rho_1 g \left(b - \frac{e^{mb}}{m} \cos m(a+ct)\right).
\end{equation}
This means that we must require
\begin{equation}
\label{densities}  \nonumber
g \rho_1 = \rho_2 (g - c^2 m),
\end{equation}
which is equivalent to 
\begin{equation}  \nonumber
c^2 m = g \frac{\rho_2 - \rho_1}{\rho_2},
\end{equation}
the right-hand side of which is simply the reduced gravity, which we denote by $g_0$, implying that the interfacial waves propagate with a celerity 
\[ c = \sqrt{g_0/m}.\]
This is clearly consistent with the propagation speed of classical Gerstner waves, which is recovered by setting $\rho_1 = 0.$
The equation of mass conservation \eqref{masscons} is equivalent to the determinant of the coordinate transform \eqref{coord_transform} being time independent, which is readily verified.

The parameter $b_0$ which defines the interface also determines the form thereof: $b_0 = 0$ corresponding to a cycloid with sharp crests, while $b_0 < 0$ describes a smooth, trochoidal profile. The assumption that $b_0 \leq 0$ is made to ensure that there are no self-intersecting particle paths. The matching of pressures \eqref{pressure_gerstner} and \eqref{pressure_still} at $b = b_0$ further necessitates that $b_0$ be the solution to the equation 
\[ \rho_2 g_0 \left( \frac{1}{2m} e^{2mb} - b\right) = P_{atm} -C. \]
This solution is unique for a suitable choice of constants, as the left-hand side is a strictly decreasing function of $b$ whose value in the limit $b \rightarrow -\infty$ is clearly $+\infty.$

\section{Discussion}
We have provided a solution describing waves with a trochoidal profile propagating at the interface between water of two different densities with a flat surface.
The Gerstner wave which forms the cornerstone of this study of dead-water belongs to the early history of water-wave theory, first discovered by its namesake in 1804 \citep{ANDP:ANDP18090320808}, and rediscovered by William J.\ M.\ Rankine more than a half-century later \citep{rankine1863exact}. A modern treatment of the Gerstner flow, proving its dynamic feasibility, was given by \citet{Constantin:2001zl} and \citet{ISI:000263517200008}. The particle motion is along circular trajectories with angular velocity $mc,$ the radius decreasing exponentially with depth. For positive celerity $c$ the waves move in the negative $x$ direction, and there is a negative vorticity associated with the wave motion -- this latter indicating that the waves cannot be generated from rest by potential forces in an inviscid, homogeneous fluid. Moreover, the circular particle paths are a hallmark of vorticity, since for irrotational Stokes waves the particle trajectories have quite a different pattern cf.\ recent work by \citet{IM}, \citet{Constantin:rr}, and \citet{Henry2010}.

It should be noted that the peculiar structure of the Gerstner solution is, perhaps, a mixed blessing -- while it furnishes us with an explicit solution whose profile is more realistic than the sinisoid of the linear theory, at the same time the nonlinear boundary conditions and governing equations are very exacting, leaving only little leeway to include other effects. At the same time, as the Gerstner wave is the only available explicit solution to the full water-wave problem, it has been modified to describe a variety of physical situations, (see for example recent work by Constantin on geophysical waves \cite{Constantin2013} and references therein). Despite this special structure, recent work on deep-water waves by \citet{Monismith2007} \emph{et al} finds that Gerstner wave theory fits observed mean velocities better than classical deep-water Stokes theory, both in the laboratory and in the ocean.

In contrast to the classical Gerstner wave solution, with the inclusion of an upper layer of still water the pressure \eqref{pressure_gerstner} is no longer constant along streamlines, but exhibits an oscillatory time-dependence. On account of this, our solution is also distinct from the Gerstner wave in stratified water \citep{0005.22804, Stuhlmeier2011}, where lines of constant density and constant pressure coincide with the streamlines. In fact, upon passing to a reference frame moving with speed $c,$ constancy of the density along streamlines, i.e.\ $\rho_a = 0,$ is ensured by the continuity equation -- in our setting, the form of the pressure precludes the possibility of exploiting this to describe a heterogeneous bottom layer.\\

The author would like to acknowledge support from ERC Grant NWFV -- Nonlinear studies of water flows with vorticity.

\end{document}